 \documentclass[aps,prl,twocolumn,superscriptaddress,longbibliography]{revtex4-2} 
\usepackage[utf8]{inputenc}
\usepackage{bbm}
\usepackage{xcolor}
\usepackage{amsmath}
\usepackage{graphicx}
\usepackage[most]{tcolorbox}
\usepackage{booktabs}
\usepackage[mathlines]{lineno}

\usepackage[colorlinks,linkcolor=blue,hyperindex,CJKbookmarks]{hyperref}

\begin{document}

\title{Evolution from three-dimensional charge density wave to one-dimensional stripe order in CsV$_{3-x}$Ti$_x$Sb$_5$}

\author{Qian Xiao}\thanks{These authors contributed equally to this work.}
\email{xq1997@pku.edu.cn}
\affiliation{School of Physics and Information Engineering, Guangdong University of Education, Guangzhou 510303, China}
\affiliation{International Center for Quantum Materials, School of Physics, Peking University, Beijing 100871, China}

\author{Xiangqi Liu}\thanks{These authors contributed equally to this work.}
\affiliation{State Key Laboratory of Quantum Functional Materials, School of Physical Science and Technology, ShanghaiTech University, Shanghai 201210, China}
\affiliation{ShanghaiTech Laboratory for Topological Physics, ShanghaiTech University, Shanghai 201210, China}

\author{Zihao Huang}\thanks{These authors contributed equally to this work.}
\affiliation{Beijing National Center for Condensed Matter Physics and Institute of Physics, Chinese Academy of Sciences, Beijing 100190, China.}
\affiliation{School of Physical Sciences, University of Chinese Academy of Sciences, Beijing 100190, China.}

\author{Xiquan Zheng}
\affiliation{International Center for Quantum Materials, School of Physics, Peking University, Beijing 100871, China}

\author{Shilong Zhang}
\affiliation{International Center for Quantum Materials, School of Physics, Peking University, Beijing 100871, China}

\author{Hui Chen}
\author{Hong-Jun Gao}
\affiliation{Beijing National Center for Condensed Matter Physics and Institute of Physics, Chinese Academy of Sciences, Beijing 100190, China.}
\affiliation{School of Physical Sciences, University of Chinese Academy of Sciences, Beijing 100190, China.}

\author{Yanfeng Guo}
\affiliation{State Key Laboratory of Quantum Functional Materials, School of Physical Science and Technology, ShanghaiTech University, Shanghai 201210, China}
\affiliation{ShanghaiTech Laboratory for Topological Physics, ShanghaiTech University, Shanghai 201210, China}

\author{Yingying Peng }
\email{yingying.peng@pku.edu.cn}
\affiliation{International Center for Quantum Materials, School of Physics, Peking University, Beijing 100871, China}
\affiliation{Collaborative Innovation Center of Quantum Matter, Beijing 100871, China}

\date{\today}

\begin{abstract}
Understanding intertwined phases near quantum criticality is a central challenge in correlated electron systems. The kagome metal CsV$_{3-x}$Ti$_x$Sb$_5$ provides a fertile platform to investigate the interplay between charge-density-wave (CDW) and superconductivity. Here, combining x-ray diffraction (XRD) and scanning tunneling microscopy (STM), we uncover a dimensional evolution of the CDW upon Ti substitution. We find that even infinitesimal Ti doping ($x = 0.009$) completely suppresses the three-dimensional $2\times2\times4$ CDW present in pristine CsV$_3$Sb$_5$, while reducing the remaining $2\times2\times2$ CDW to a quasi-two-dimensional order. With further Ti substitution, although no CDW transition is discernible in resistivity measurements, our XRD and STM data reveal the emergence of a (quasi-)one-dimensional CDW with a short correlation length of $\sim20$~\AA\ at $x = 0.2$. The stripe-like CDW undergoes a continuous second-order phase transition, characterized by a gradual increase in intensity and correlation length below $\sim56$~K. Our results elucidate the dimensional evolution of CDW order in CsV$_{3-x}$Ti$_x$Sb$_5$ and provide new insight into understanding the unconventional
CDWs and their role in kagome superconductors.

\end{abstract}

\maketitle
The phase landscape of quantum materials is often governed by the intricate competition and coexistence of multiple electronic orders\,\cite{Lee.RMP.2006,PNAS.intertwined,cuprate.RevModPhys.87.457}. The generic emergence of superconductivity upon the suppression of a primary long-range order, such as antiferromagnetism or charge density waves (CDWs), highlights the pivotal role of the underlying fluctuating regime\,\cite{BKeimer.review.nat.2015,Fernandes.nat,Riccardo.2021.review.chargeOrder.cuprate}. Elucidating the nature of the precursor states, exemplified by the enigmatic pseudogap phase in cuprates\,\cite{BKeimer.review.nat.2015}, the intertwined nematic and magnetic fluctuations in iron-based superconductors\,\cite{Fernandes.nat,MiaoQiSi.2016.nat.rev.mater}, and the proposed quantum critical regimes in heavy-fermion systems\,\cite{Gegenwart.2008.natphy.heavyFermiMetal}, remains a central challenge in condensed matter physics.

The kagome metal CsV$_3$Sb$_5$ exhibits a wealth of emergent phenomena and provides an exciting platform for investigating competing and intertwined electronic orders. Its prototypical kagome lattice naturally hosts nontrivial electronic structures, including Dirac crossings, van Hove singularities (vHSs), and flat bands, thereby promoting a rich interplay among electronic correlations, topology, and symmetry breaking\,\cite{prm2019Brenden,Brenden.prl,cpl2021RVS.sc,jiangyx}. Pristine CsV$_3$Sb$_5$ undergoes a CDW transition below $\sim94$~K and becomes superconducting below $\sim3$~K. Notably, coexisting and competing $2\times2\times2$ and $2\times2\times4$ CDW phases have been identified, originating from distinct interlayer stacking configurations of in-plane inverse star-of-David distortions within the kagome layers\,\cite{xq,sdH.prl.CVS}. In addition, a long-range $1\times4$ ($4a$) charge order has been observed by scanning tunneling microscopy (STM)\,\cite{HeZhao.nat.2021}. Intriguingly, the CDW phase in CsV$_3$Sb$_5$ has been proposed to be chiral and to exhibit time-reversal symmetry breaking (TRSB)\,\cite{AHE.CVS,umR.kenny,miuSR.nat,nematicity.3domains.natphy}, and to intertwine with electronic nematicity\,\cite{xq,lihong.natphy,nematicity.CsVTiSb.natphy,Nie.2022.nat,NLWang.TRS}.

External tuning by pressure or chemical substitution reveals an unconventional competition between CDW order and superconductivity in CsV$_3$Sb$_5$\,\cite{LixuanZheng.nat.pressure.CVS,F.H.Yu.pressure.nc,CVSSn.xrd.stripe,CVTiSb.STM,CVTiSb.PRL.2025.QPI,Zhong.CVNS.CVTS.nat,CVNS.ARPES.PRL}. In particular, transport measurements show that Ti substitution initially suppresses both the superconducting transition temperature $T_{\mathrm{c}}$ and the CDW transition temperature $T_{\mathrm{CDW}}$, while $T_{\mathrm{c}}$ increases after complete suppression of CDW at higher Ti concentrations\,\cite{CVTiSb.STM}. Although no CDW transition is discernible in resistivity measurements for heavily Ti-doped samples, STM studies have revealed the persistence of a gaplike feature in CsV$_{3-x}$Ti$_x$Sb$_5$ ($x = 0.15$)\,\cite{CVTiSb.PRL.2025.QPI}, suggesting the presence of a hidden or unconventional electronic order in the highly doped regime. However, direct x-ray scattering evidence for CDW order in highly Ti-doped CsV$_3$Sb$_5$ is still lacking.

In this work, through systematic x-ray diffraction and scanning tunneling microscopy studies, we reveal an evolution from the three-dimensional charge density wave to the one-dimensional stripe order in the kagome superconductor CsV$_{3-x}$Ti$_x$Sb$_5$. We find that even an infinitesimal Ti substitution ($x = 0.009$) completely suppresses the three-dimensional $2\times2\times4$ CDW and drives the remaining $2\times2\times2$ order into a quasi-two-dimensional state. 
In the heavily Ti-doped compounds, the CDW phase further evolves into a quasi-1D CDW state characterized by short-range stripe charge correlations. These stripe correlations undergo a second-order phase transition at a critical temperature of $\sim$ 56\,K, characterized by a gradual enhancement of both the diffraction intensity and the correlation length upon cooling. Our results elucidate the microscopic dimensional evolution of the CDW order in CsV$_{3-x}$Ti$_x$Sb$_5$ and establish this system as a versatile platform for investigating the interplay between dimensionality, stripe order, and superconductivity.

Single crystals of CsV$_{3-x}$Ti$_x$Sb$_5$ were grown using a self-flux method. The chemical composition was characterized by scanning electron microscopy (SEM) equipped with energy-dispersive x-ray spectroscopy (EDX). The Ti concentration was determined by averaging measurements taken at multiple positions on each crystal, yielding an uncertainty of approximately 1$\%$.
Single-crystal x-ray diffraction measurements were carried out on a custom-designed diffractometer equipped with a Xenocs Genix3D Mo K$_\alpha$ x-ray source (17.48 keV)\,\cite{xq}. In pristine CsV$_3$Sb$_5$, the lattice parameters are $a=b\simeq5.53$~\AA\ and $c\simeq9.28$~\AA. Upon Ti substitution for V, all lattice parameters decrease monotonically, consistent with the smaller ionic radius of Ti compared to V. For CsV$_{3-x}$Ti$_x$Sb$_5$ with $x=0.2$, the lattice parameters are reduced to $a=b=5.52$~\AA\ and $c=9.26$~\AA.
Throughout this manuscript, the Miller indices $(H,K,L)$ are defined with respect to the high-temperature crystal structure; details of the indexing scheme are provided in Ref.~\cite{xq}.

\begin{figure}[htbp]
\centering
\includegraphics[width=\linewidth]{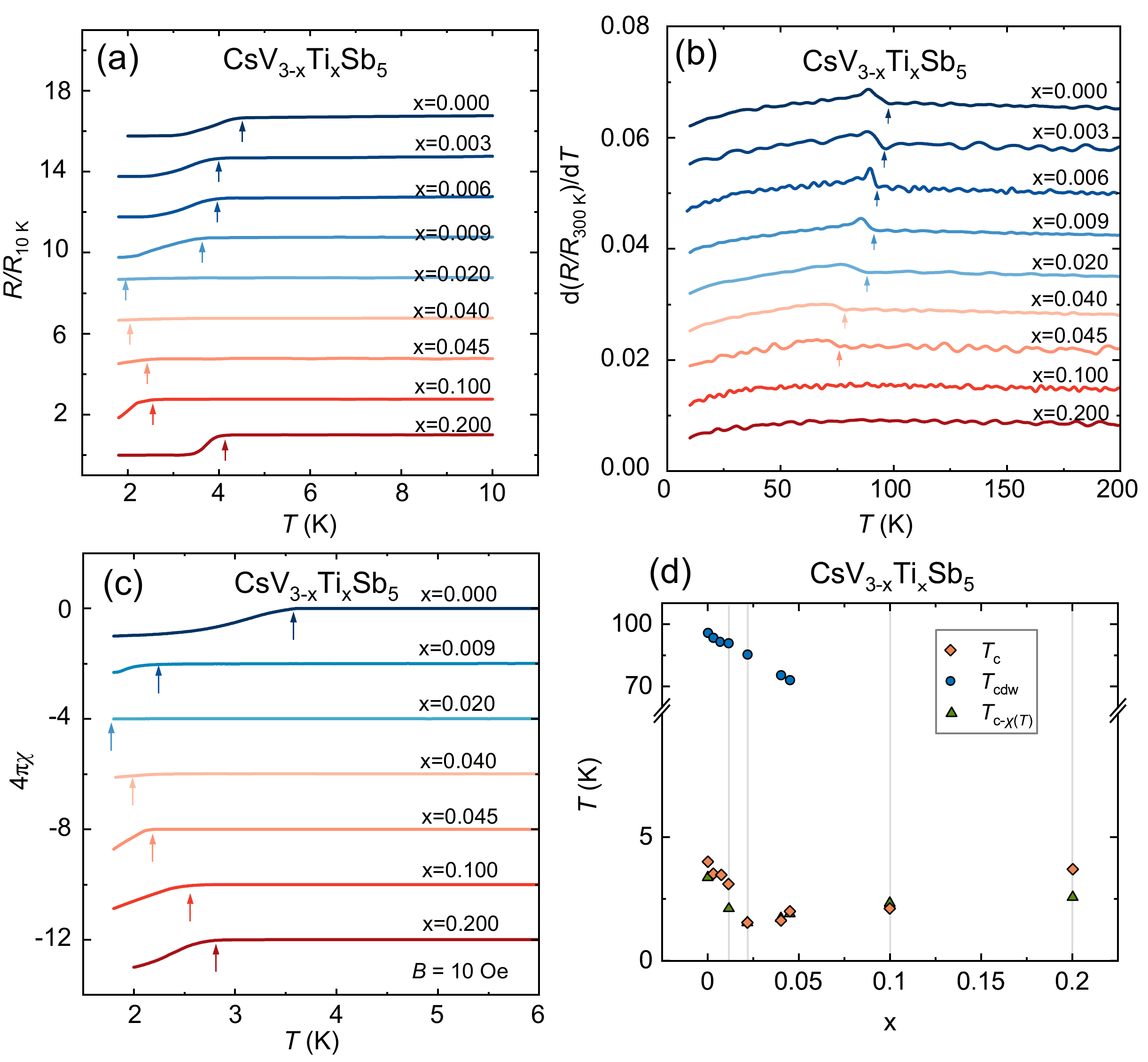}\caption{Superconductivity and CDW in ${\mathrm{CsV}}_{3-x}{\mathrm{Ti}}_{x}{\mathrm{Sb}}_5$. (a) Temperature dependence of the in-plane normalized resistance R/R$_{300 K}$. The superconducting transition temperature $T_{\mathrm{c}}$ is determined by the onset temperature of resistance drop. (b) The derivative electrical resistance dR/dT as a function of temperature. The CDW transition temperature $T_{\mathrm{CDW}}$ was determined by the onset point of dR/dT jump. (c) Temperature dependent of magnetic susceptibilities under zero-field cooling (ZFC) with applied field of 10 Oe along c-axis. The data is corrected by calculating the demagnetization factor. Offset was added for clarity. (d) Phase diagram of ${\mathrm{CsV}}_{3-x}{\mathrm{Ti}}_{x}{\mathrm{Sb}}_5$ crystals. The solid lines indicate the selected Ti doping level in our XRD study.
\label{fig1}}
\end{figure}

Figure~\ref{fig1} shows the resistance and magnetic susceptibility measurements of CsV$_{3-x}$Ti$_x$Sb$_5$ single crystals. Superconductivity is evidenced by a sharp drop in the resistance and the onset of diamagnetism, as shown in Figs.~\ref{fig1}(a) and \ref{fig1}(c), respectively. A pronounced anomaly in the temperature derivative of resistance, $dR/dT$ (Fig.~\ref{fig1}(b)), signals the CDW phase transition.
With increasing Ti substitution, the CDW transition is progressively suppressed: $T_{\mathrm{CDW}}$ decreases and the associated anomaly in $dR/dT$ becomes increasingly broadened and weakened up to $x=0.045$. For higher Ti concentrations ($x>0.045$), no discernible CDW transition can be resolved in the transport data. The resulting phase diagram of CsV$_{3-x}$Ti$_x$Sb$_5$ is summarized in Fig.~\ref{fig1}(d).
The superconducting transition temperature $T_{\mathrm{c}}$ exhibits a non-monotonic evolution with Ti doping. It is initially suppressed upon Ti substitution up to $x\approx0.04$, followed by a recovery and enhancement at higher doping levels, in agreement with previous reports,\cite{nematicity.CsVTiSb.natphy}.

\begin{figure}[htbp]
\centering
\includegraphics[width=\linewidth]{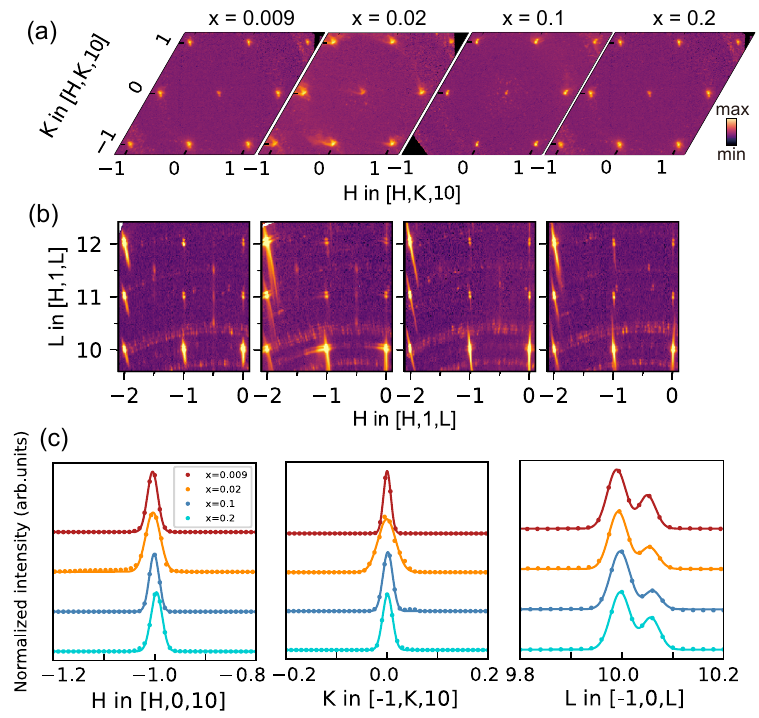}\caption{XRD measurements of ${\mathrm{CsV}}_{3-x}{\mathrm{Ti}}_{x}{\mathrm{Sb}}_5$ taken at 18 K for four Ti doping levels using a Mo K$_{\alpha}$ x-ray source. (a) $(H, K)$ maps of reciprocal space at $L$ = 10. (b) $(H, L)$ maps of reciprocal space at $K$ = 1. (c) $H$-, $K$- and $L$-cuts of the main Bragg peak at [-1, 0, 10] are shown in the left, middle and right panel, respectively. Intensities are normalized to maximum. Vertical offset are added for clarity. The double peaks in $L$-cut originate
from an x-ray source consisting of Mo $K_{\alpha1}$ and $K_{\alpha2}$\,\cite{xq}. The solid lines are fits with Gaussian functions. 
\label{fig2-2}}
\end{figure}

To elucidate the impact of Ti substitution on the CDW phases in this kagome system, we performed systematic XRD measurements on CsV$_{3-x}$Ti$_x$Sb$_5$ single crystals spanning representative regions of the phase diagram. Four doping levels were selected: $x=0.009$, located within the first superconducting dome; $x=0.02$, near the boundary separating the two superconducting domes; and $x=0.1$ and $0.2$, located inside the second superconducting dome, where no discernible CDW anomalies are observed in the transport measurements.
XRD measurements conducted at 18~K (Fig.~\ref{fig2-2}) reveal sharp and intense fundamental Bragg reflections for all four compositions, attesting to their high crystalline quality. Among them, the $x=0.02$ sample exhibits slightly broadened Bragg peaks, indicating a modest reduction in crystalline perfection compared to the other doped crystals.

\begin{figure}[htbp]
\centering
\includegraphics[width=\linewidth]{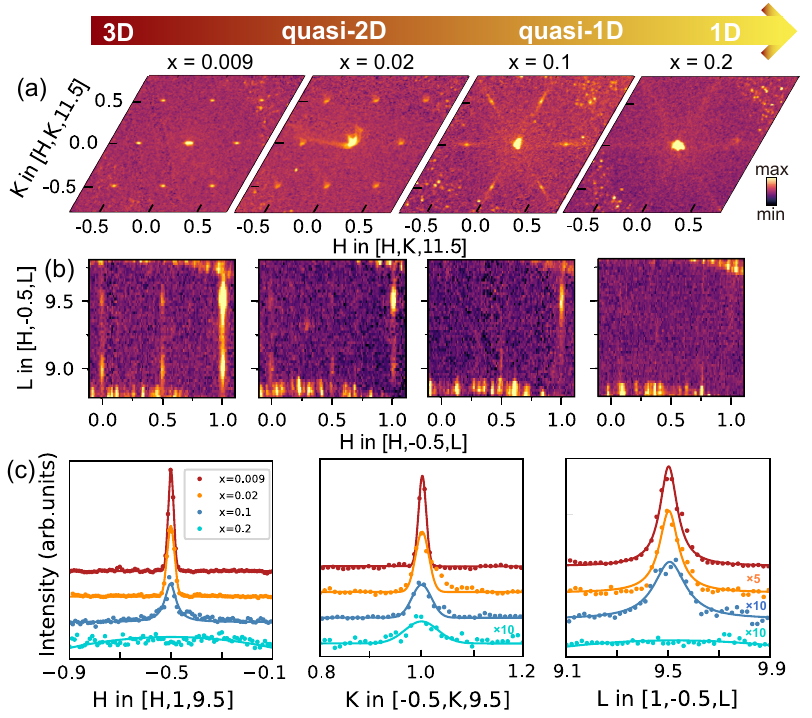}\caption{XRD measurements of ${\mathrm{CsV}}_{3-x}{\mathrm{Ti}}_{x}{\mathrm{Sb}}_5$ taken at 18 K for four Ti doping levels using a Mo K$_{\alpha}$ x-ray source. (a) $(H, K)$ maps of reciprocal space at $L$ = 11.5. The diffraction point at the center is the leakage from main Bragg peak. (b) $(H, L)$ maps of reciprocal space at $K$ = -0.5. The arc signals in colormaps come from beryllium domes. (c) $H$-, $K$-cuts of [-0.5, 1, 9.5] and $L$-cuts of [1, -0.5, 9.5] are shown in the left, middle and right panel, respectively. Vertical offsets are applied for clarity. $H$-cuts: Gaussian fits for $x = 0.009$ and $0.02$, Lorentzian fit for $x = 0.1$, and a guide-to-the-eye line for $x = 0.2$. $K$-cuts: Gaussian fits for all doping levels. $L$-cuts: Lorentzian fits for $x$ = 0.009, 0.02 and 0.1, and a guide-to-the-eye line for $x$ = 0.2. In the fitting, Lorentz functions are used to capture the stripe-like feature in line profile\,\cite{ScV6Sn6.diffuse.scattering}.
\label{fig2}}
\end{figure}

Figures~\ref{fig2}(a, b) present the evolution of the CDW diffraction patterns at 18~K as a function of Ti doping. All samples were cooled to 18~K at a rate of 2~K/min prior to measurement. Already at the lowest doping level ($x=0.009$), the CDW reflections exhibit pronounced broadening along the out-of-plane direction in the $(H,L)$ reciprocal-space maps, whereas well-defined and discrete CDW peaks persist in the $(H,K)$ planes up to $x=0.02$. This marked anisotropy demonstrates that even infinitesimal Ti substitution strongly suppresses the interlayer CDW correlations, effectively driving the CDW toward a quasi-two-dimensional state.
Figure~\ref{fig2}(c) summarizes the doping dependence of representative $H$-, $K$-, and $L$-direction cuts through the CDW peaks. For $x=0.009$, the in-plane correlation length $\xi_K$, defined as the inverse of the half width at half maximum (HWHM) obtained from peak fitting, is reduced by approximately 33\% compared to the pristine compound (from $\sim127$~\AA\ to $\sim85$~\AA). In contrast, the out-of-plane correlation length $\xi_L$ is reduced by about 68\% (from $\sim114$~\AA\ to $\sim36$~\AA), highlighting the stronger destruction of interlayer correlation by Ti doping.  Furthermore, Ti substitution completely destabilizes the $2\times2\times4$ CDW phase, as evidenced by the absence of quarter-integer $L$-position superlattice peaks and the disappearance of thermal hysteresis in all Ti-doped samples (discussed below), consistent with earlier observations in Ta- and Nb-substituted systems\,\cite{xq.Ta.Nb}.

In the higher Ti-doping regime ($x=0.1$ and $0.2$), corresponding to the second superconducting dome where transport measurements reveal no discernible CDW transition, our XRD measurements detect clear CDW scattering signals. Notably, these signals exhibit pronounced stripe-like features in both the $(H,K)$ and $(H,L)$ reciprocal-space maps, indicating that the CDW evolves into a quasi-one-dimensional stripe order. For $x=0.2$, the CDW reflections are significantly weakened and broadened, forming diffuse rod-like features extending along both in-plane ($H$ or $K$) and out-of-plane ($L$) directions. Transverse to the rod direction, the extracted correlation length is approximately 20~\AA\ ($\sim4a$), reflecting strongly enhanced anisotropy and short-range stripe correlations. Importantly, this dimensional reduction of the CDW with increasing Ti content cannot be attributed to deteriorating crystal quality, as the $x=0.1$ and $0.2$ samples exhibit superior crystallinity compared to the $x=0.02$ sample, as demonstrated in Fig.~\ref{fig2-2}.

\begin{figure}[htbp]
\centering
\includegraphics[width=0.95\linewidth]{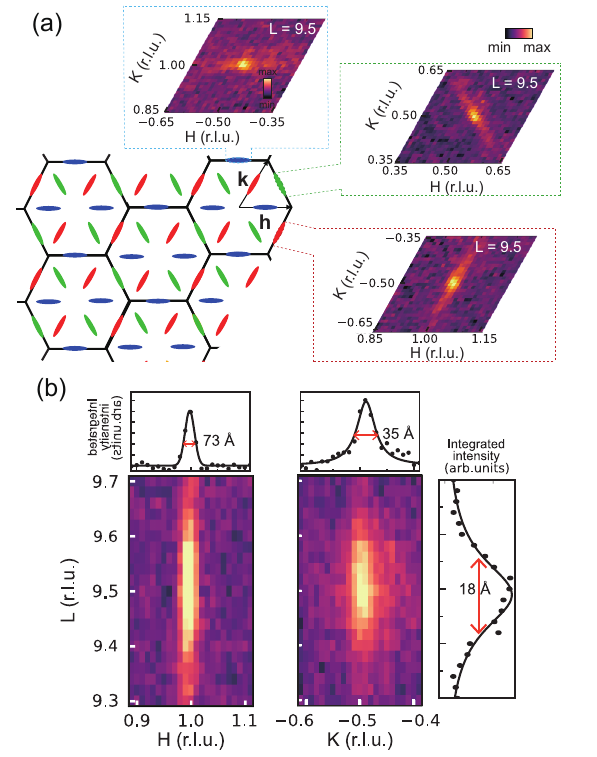}\caption{XRD measurements for ${\mathrm{CsV}}_{3-x}{\mathrm{Ti}}_{x}{\mathrm{Sb}}_5$ with x = 0.1 at 18 K using a Mo K$_{\alpha}$ x-ray source. 
(a) Schematic of scattering signals in $(H, K)$ maps. Different colors indicate different orientations of electronic correlations. Three representative experimental $(H, K)$ diffraction maps are shown in dotted rectangular box. (b) $(H, L)$ and $(K, L)$ maps at $\bf {q}$ = [1, -0.5, 9.5]. The corresponding $H$-, $K$- and $L$-cut are shown near their colormap, respectively. Solid lines in $H$-cut and $K/L$-cut are fits with Gaussian and Lorentz functions, respectively. The intensity of colormaps was plotted in log scale.
\label{fig3}}
\end{figure}

\begin{figure}[htbp]
\centering
\includegraphics[width=\linewidth]{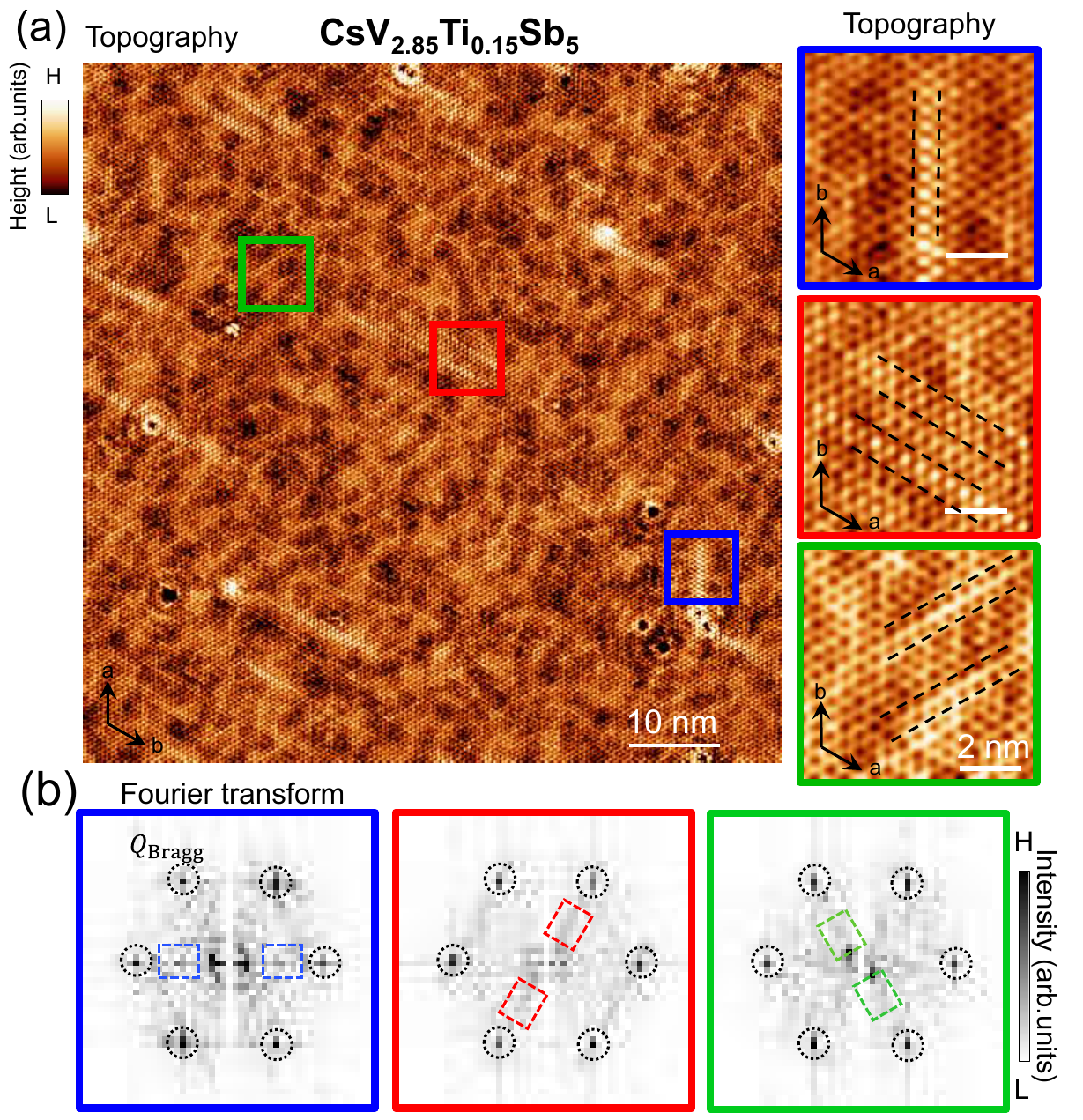}\caption{Observations of three equivalent domains of short-ranged unidirectional charge orders in CsV$_{2.85}$Ti$_{0.15}$Sb$_5$ crystal. (a) STM images showing short-ranged charge orders with multiple nanoscale domains. Right panel: zoom-in images of the areas outlined by green, red, and blue squares in the overview image on the left.
Each magnified region displays a dominant unidirectional charge order, aligned predominantly along one of three distinct crystalline directions, as indicated by the black dashed lines. (b) Fourier transform of the zoom-in images in (a), showing the broad peaks around the 1/2 {\bf{Q}}$_{\mathrm{Bragg}}$ wave vector along one of the three crystalline directions (marked by colored dashed square). Scanning parameters: bias volage Vs=-50 mV, tunneling current: $I_{\mathrm{t}}$=1\,nA; temperature: 0.4\,K. 
\label{fig3-3}}
\end{figure}

These stripes are centered on the same $(H, K, L)$ as that of the 2\,$\times$2\,$\times$2 CDW phase in the pristine sample, suggesting that the CDW is preserved in the Ti-doped sample.
However, the correlation length along the out-of-plane direction is only about 18\,\AA, i.e., 2 unit cells, indicating that the phase shift of adjacent kagome layers exists only between neighbor layers in the sample $x = 0.1$ \,\cite{jiangyx,sdh2021Brenden,xq}. 
These short-range correlations are clearly anisotropic
in the plane. 
As shown in Fig.\,\ref{fig3}, at ${\bf q}$ = [1, -0.5, 9.5] in $x = 0.1$ sample, the in-plane correlation lengths $\xi$ along $H$ and $K$ direction are $\sim$ 73\,\AA\ and 35\,\AA, respectively.
The estimated in-plane anisotropy, defined as ($\xi_H - \xi_K$)/($\xi_H + \xi_K$), is approximately 35\%. 
As shown in Fig.\,\ref{fig3}(a), three distinct preferential orientations are observed in their reciprocal-space ($H, K$) plane.
The in-plane short axis of the stripes is found to be orthogonal to the ${\bf{h}}$/${\bf{k}}$ or parallel to ${\bf{h}}$+${\bf{k}}$ direction.
This indicates that charge correlations develop preferentially along a single in-plane axis, e.g., the lattice axis $\vec{a}$ or $\vec{b}$ or $\vec{a}+\vec{b}$ direction, breaking the sixfold rotational symmetry down to twofold, and leading to the formation of three distinct real-space domains oriented at 120$^{\circ}$ relative to one another \,\cite{nematicity.3domains.natphy}. 
This is made possible in CsV$_3$Sb$_5$ by the presence of three equivalent stacking orientations between adjacent kagome layers\,\cite{xq,mattieu.prl.pressure.cvs}.
The two coupled kagome layer are organized as stripe-like structures in the sample $x = 0.1$, which align with the stacking directions of adjacent kagome layers. This scenario is confirmed by the morphology of the real-space electronic signal. 
The STM topography of $x = 0.15$ Ti-doped sample reveals three equivalent nanoscale domains of short-range unidirectional charge order, each predominantly aligned along one of the three distinct crystalline directions (Fig.\,\ref{fig3-3}).

\begin{figure}[htbp]
\centering
\includegraphics[width=\linewidth]{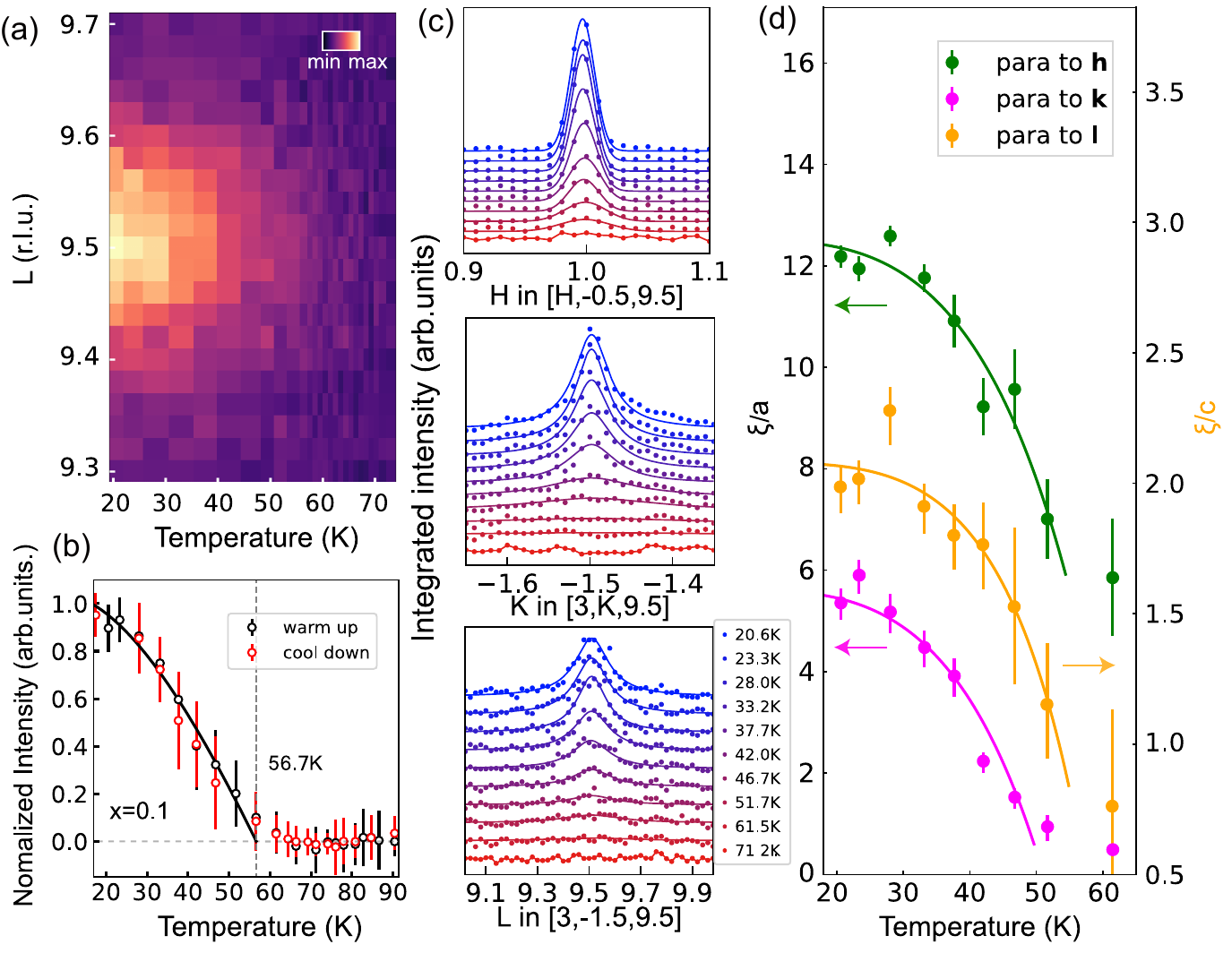}
\caption{Momentum scans and correlation lengths at different
temperatures for CsV$_{3-x}$Ti$_x$Sb$_5$   with x = 0.1. (a) Colormap of reciprocal space at [1, -0.5, 9.5] versus temperature. (b) The normalized integrated intensity versus temperature. The data were obtained by averaging the scattering signals of more than ten CDW peaks, error bars came from the deviation from the average value. Solid line is fit with a mean-field function. (c) The waterfalls of $H$-, $K$- and $L$-cuts at several temperatures. Solid lines in $H$-cuts are fits with Gaussian functions. Solid lines in $K$-cuts and $L$-cuts are fits with Lorentz functions. The extracted correlation lengths varying as temperature are shown in (d). Error bars in (d) were the standard deviation of the fits. Solid lines are guide for eyes.
\label{fig4}}
\end{figure}

We next examine the temperature evolution of the stripe correlations in CsV$_{3-x}$Ti$_x$Sb$_5$ with $x=0.1$, as shown in Fig.~\ref{fig4}. No thermal hysteresis is observed between cooling and warming cycles, confirming that the first-order $2\times2\times4$ CDW phase is completely suppressed in this composition. In stark contrast to pristine CsV$_3$Sb$_5$, the stripe CDW intensity decreases smoothly with increasing temperature, indicating that the transition is continuous and consistent with a second-order phase transition.
The CDW transition temperature for the $x=0.1$ sample is estimated to be $T_{\mathrm{CDW}} = 56.7 \pm 2.3$~K by fitting the temperature-dependent stripe intensity to a mean-field form $I(T) = A\,\mathrm{tanh}^2\bigg(C\sqrt{\frac{T{_{\mathrm{CDW}}}}{T}-1}\bigg)$, 
where $A$ and $C$ are fitting parameters\,\cite{singleTiSe2.nc.CDW,TiTeSe.nano.CDW}. Figures~\ref{fig4}(c) and \ref{fig4}(d) display representative temperature-dependent CDW peak profiles along the $H$, $K$, and $L$ directions and the corresponding extracted correlation lengths. Upon cooling below $T_{\mathrm{CDW}}$, the correlation lengths increase progressively and exhibit pronounced anisotropy along the three crystallographic directions, reflecting the intrinsically low-dimensional nature of the CDW correlations in this doping regime.

Our results demonstrate that Ti substitution on the V sites progressively suppresses the CDW correlation length and drives a dimensional crossover of the CDW order, evolving from long-range three-dimensional coherence to short-range, stripe-like correlations. In heavily Ti-doped CsV$_{3-x}$Ti$_x$Sb$_5$, the absence of a discernible CDW transition in transport measurements signals the collapse of long-range order. Instead, the system retains short-range stripe correlations that preserve local symmetry breaking. This picture is consistent with recent $^{51}$V nuclear magnetic resonance (NMR) measurements, which revealed the emergence of a unidirectional CDW localized around Ti dopants for doping levels above $x=0.12$\,\cite{2025.NMR.PRB.CsV3-xTixSb5}. These findings support an intrinsic stripe character of the CDW in this doping regime. Moreover, the presence of stripe CDW order naturally accounts for the gaplike features previously observed in STM measurements on highly Ti-doped samples ($x=0.15$)\,\cite{CVTiSb.PRL.2025.QPI}.

Notably, the short-range stripe correlations uncovered here are distinct from those reported in Sn-substituted CsV$_3$Sb$_{5-x}$Sn$_x$, where heavily Sn-doped samples develop stripe order with an in-plane wavevector ${\bf q}{_\mathrm{in\text{-}plane}} \sim 1/3$\,\cite{cxh2025.prx.CsV3Sb5-xSnx.stripe,CVSSn.xrd.stripe}. Although both Ti and Sn substitution introduce hole doping, the different substitution sites, V versus Sb sublattices, lead to different microscopic mechanisms. In CsV$_3$Sb$_{5-x}$Sn$_x$, the stripe order at ${\bf q}{_\mathrm{in\text{-}plane}} \sim 1/3$ is primarily driven by strong in-plane chemical pressure induced by Sn substitution. This chemical pressure shifts the dominant lattice instability toward the $1/3$ wavevector, with nonlocal electronic correlations near the van Hove singularities playing a key role in stabilizing the stripe phase\,\cite{cxh2025.prx.CsV3Sb5-xSnx.stripe}.
In contrast, Ti substitution introduces three intertwined perturbations: hole doping, orbital selectivity, and disorder. Hole doping shifts the Fermi level away from the van Hove singularity, thereby thermodynamically weakening the electronic instability responsible for long-range CDW order. Concurrently, Ti substitution on the V sites induces orbital-selective scattering, which suppresses the out-of-plane coherence of the CDW and hinders the formation of three-dimensional order\,\cite{CVTiSb.PRL.2025.QPI}. Finally, the disorder introduced by Ti dopants kinetically disrupts phase coherence and acts as a random pinning landscape toward unidirectional CDW\,\cite{2025.NMR.PRB.CsV3-xTixSb5}. 

The short-range stripe order observed in heavily Ti-doped CsV$_3$Sb$_5$ is reminiscent of the anisotropic CDW fluctuations reported in 1T-TiSe$_2$, where diffuse scattering is detected well above $T{_\mathrm{CDW}}$\,\cite{guo2025TiSe2}. However, the stripe correlations in CsV$_{3-x}$Ti$_x$Sb$_5$ develop only below $\sim 56$ K and do not exhibit a divergence of the correlation length upon approaching the CDW transition. This behavior indicates that the stripe state is not a simple manifestation of the critical CDW fluctuations, but instead represents a distinct short-range ordered regime. 
Nevertheless, the unidirectional nature of the stripe correlations implies the presence of strong nematic fluctuations. Such nematicity has been widely observed in CsV$_3$Sb$_5$–based systems\,\cite{XiangYing.nc.2fold.magnetoResi,Nie.2022.nat,nematicity.3domains.natphy,nematicity.CsVTiSb.natphy} and is found to persist, together with charge fluctuations, deep into the doping regime associated with the second superconducting dome. This observation is consistent with recent reports of persistent CDW dynamics in doped kagome systems\,\cite{2025.arxiv.persist.CDW}. More broadly, a fluctuating electronic background of this type is often regarded as favorable for unconventional superconductivity, suggesting a potential connection between stripe-like charge correlations and superconductivity in kagome metals\,\cite{Fernandes.nat,E.Fradkin2015RMP,FeSeS.PNAS.2016.namtic}, which warrants further investigation.

In summary, we demonstrate that Ti doping in the kagome metal CsV$_3$Sb$_5$ induces a collapse of the three-dimensionally long-range CDW into a short-range stripe order. Remarkably, the short-range CDW phase in heavily Ti-doped samples undergoes a second-order phase transition, in contrast to the first-order CDW transition observed in pristine CsV$_3$Sb$_5$. This stripe-dominated regime coincides with the emergence of the second superconducting dome, pointing to a close interplay between unidirectional charge order and superconductivity, and motivating further experimental and theoretical investigations.

We acknowledge the valuable discussions with Chandra Varma, Steven Kivelson, Geng Li and Zhenyu Wang. Y.~Y.~P. is grateful for financial support from the National Natural Science Foundation of China (Grant No. 12374143), the Ministry of Science and Technology of China (Grants No. 2021YFA1401903 and No. 2024YFA1408702), and the Beijing Natural Science Foundation
(Grant No. JQ24001). Y.~F.~G.’s work
is supported by the Science and Technology Commission of Shanghai Municipality (Grant No. 25DZ3008200). H.~J.~G. and H.~C. acknowledge the financial
support from the National Natural Science Foundation of China (Grant No. 62488201), the National Key Research and Development Projects of China (Grant No. 2022YFA1204100).

\bibliographystyle{apsrev4-2}
\bibliography{reference}

@article{xq,
  title = "{Coexistence of multiple stacking charge density waves in kagome superconductor ${\mathrm{CsV}}_{3}{\mathrm{Sb}}_{5}$}",
  author = {Xiao, Qian and Lin, Yihao and Li, Qizhi and Zheng, Xiquan and Francoual, Sonia and Plueckthun, Christian and Xia, Wei and Qiu, Qingzheng and Zhang, Shilong and Guo, Yanfeng and others},
  journal = {Phys. Rev. Res.},
  volume = {5},
  issue = {1},
  pages = {L012032},
  numpages = {6},
  year = {2023},
  month = {Mar},
  publisher = {American Physical Society},
  doi = {10.1103/PhysRevResearch.5.L012032},
}

@article{Brenden.prl,
  title = "{$\mathrm{Cs}{\mathrm{V}}_{3}{\mathrm{Sb}}_{5}$: A ${\mathbb{Z}}_{2}$ topological Kagome metal with a superconducting ground state}",
  author = {Ortiz, Brenden R. and Teicher, Samuel M. L. and Hu, Yong and Zuo, Julia L. and Sarte, Paul M. and Schueller, Emily C. and Abeykoon, A. M. Milinda and Krogstad, Matthew J. and Rosenkranz, Stephan and Osborn, Raymond and others},
  journal = {Phys. Rev. Lett.},
  volume = {125},
  issue = {24},
  pages = {247002},
  numpages = {6},
  year = {2020},
  month = {Dec},
  publisher = {American Physical Society},
  doi = {10.1103/PhysRevLett.125.247002}
}

@article{xq.Ta.Nb,
  title = "{Evolution of charge density waves from three-dimensional to quasi-two-dimensional in kagome superconductors $\mathrm{Cs}{({\mathrm{V}}_{1\ensuremath{-}x}{M}_{x})}_{3}{\mathrm{Sb}}_{5}$ $(M=\mathrm{Nb}, \mathrm{Ta})$}",
  author = {Xiao, Qian and Li, Qizhi and Liu, Jinjin and Li, Yongkai and Xia, Wei and Zheng, Xiquan and Guo, Yanfeng and Wang, Zhiwei and Peng, Yingying},
  journal = {Phys. Rev. Mater.},
  volume = {7},
  issue = {7},
  pages = {074801},
  numpages = {10},
  year = {2023},
  month = {Jul},
  publisher = {American Physical Society},
  doi = {10.1103/PhysRevMaterials.7.074801},
  url = {https://link.aps.org/doi/10.1103/PhysRevMaterials.7.074801}
}

@article{guo2025TiSe2,
  title = "{In-Plane Anisotropy of Charge Density Wave Fluctuations in $1T\text{\ensuremath{-}}{\mathrm{TiSe}}_{2}$}",
  author = {Guo, Xuefei and Kogar, Anshul and Henke, Jans and Flicker, Felix and de Juan, Fernando and Sun, Stella X.-L. and Khayr, Issam and Peng, Yingying and Lee, Sangjun and Krogstad, Matthew J. and Rosenkranz, Stephan and Osborn, Raymond and Ruff, Jacob P. C. and Lioi, David B. and Karapetrov, Goran and Campbell, Daniel J. and Paglione, Johnpierre and van Wezel, Jasper and Chiang, Tai C. and Abbamonte, Peter},
  journal = {Phys. Rev. Lett.},
  volume = {135},
  issue = {13},
  pages = {136102},
  numpages = {6},
  year = {2025},
  month = {Sep},
  publisher = {American Physical Society},
  doi = {10.1103/j8vm-wb65},
}

@article{jiangyx,
  title = "{Unconventional chiral charge order in kagome superconductor ${\mathrm{KV}}_{3}{\mathrm{Sb}}_{5}$}",
  author = {Jiang, Yu-Xiao and Yin, Jia-Xin and Denner, M. Michael and Shumiya, Nana and Ortiz, Brenden R. and Xu, Gang and Guguchia, Zurab and He, Junyi and Hossain, Md Shafayat and Liu, Xiaoxiong and Ruff, Jacob and Kautzsch, Linus and Zhang, Songtian S. and Chang, Guoqing and Belopolski, Ilya and Zhang, Qi and Cochran, Tyler A. and Multer, Daniel and Litskevich, Maksim and Cheng, Zi-Jia and Yang, Xian P. and Wang, Ziqiang and Thomale, Ronny and Neupert, Titus and Wilson, Stephen D. and Hasan, M. Zahid},
  journal = {Nature Materials},
  volume = {20},
  issue = {10},
  pages = {1353-1357},
  year = {2021},
  month = {Oct},
  doi = {10.1038/s41563-021-01034-y}
}

@article{sdh2021Brenden,
  title = "{Fermi Surface Mapping and the Nature of Charge-Density-Wave Order in the Kagome Superconductor ${\mathrm{CsV}}_{3}{\mathrm{Sb}}_{5}$}",
  author = {Ortiz, Brenden R. and Teicher, Samuel M. L. and Kautzsch, Linus and Sarte, Paul M. and Ratcliff, Noah and Harter, John and Ruff, Jacob P. C. and Seshadri, Ram and Wilson, Stephen D.},
  journal = {Phys. Rev. X},
  volume = {11},
  issue = {4},
  pages = {041030},
  numpages = {14},
  year = {2021},
  month = {Nov},
  publisher = {American Physical Society},
  doi = {10.1103/PhysRevX.11.041030}
}

@article{CVTiSb.STM,
  title = "{Titanium doped kagome superconductor ${\mathrm{CsV}}_{3-x}{\mathrm{Ti}}_{x}{\mathrm{Sb}}_{5}$ and two distinct phases}",
  author = {Haitao Yang and Zihao Huang and Yuhang Zhang and Zhen Zhao and Jinan Shi and Hailan Luo and Lin Zhao and Guojian Qian and Hengxin Tan and Bin Hu and Ke Zhu and Zouyouwei Lu and Hua Zhang and Jianping Sun and Jinguang Cheng and Chengmin Shen and Xiao Lin and Binghai Yan and Xingjiang Zhou and Ziqiang Wang and Stephen J. Pennycook and Hui Chen and Xiaoli Dong and Wu Zhou and Hong-Jun Gao},
  journal = {Science Bulletin},
  volume = {67},
  issue = {21},
  pages = {2176},
  numpages = {10},
  year = {2022},
  month = {Nov},
  doi = {10.1016/j.scib.2022.10.015}
}

@article{2025.NMR.PRB.CsV3-xTixSb5,
  title = "{Competitive charge density waves in the doped kagome superconductor ${\mathrm{CsV}}_{3\ensuremath{-}x}{\mathrm{Ti}}_{x}{\mathrm{Sb}}_{5}$}",
  author = {Wu, Zhimian and Sun, Kuanglv and Li, Hongyu and Nie, Linpeng and Rao, Huachen and Zhao, Dan and Xiang, Ziji and Ying, Jianjun and Wang, Zhenyu and Wu, Tao and Chen, Xianhui},
  journal = {Phys. Rev. B},
  volume = {112},
  issue = {14},
  pages = {144512},
  numpages = {8},
  year = {2025},
  month = {Oct},
  publisher = {American Physical Society},
  doi = {10.1103/ls5k-8q3d}
}

@article{nematicity.3domains.natphy,
  title = "{Three-state nematicity and magneto-optical Kerr effect in the charge density waves in kagome superconductors}",
  author = {Xu, Yishuai and Ni, Zhuoliang and Liu, Yizhou and Ortiz, Brenden R. and Deng, Qinwen and Wilson, Stephen D. and Yan, Binghai and Balents, Leon and Wu, Liang},
  journal = {Nature Physics},
  volume = {18},
  issue = {12},
  pages = {1470-1475},
  year = {2022},
  month = {Dec},
  doi = {10.1038/s41567-022-01805-7}
}

@article{miuSR.nat,
  title = "{Time-reversal symmetry-breaking charge order in a kagome superconductor}",
  author = {Mielke, C. and Das, D. and Yin, J. X. and Liu, H. and Gupta, R. and Jiang, Y. X. and Medarde, M. and Wu, X. and Lei, H. C. and Chang, J. and others},
  journal = {Nature},
  year = {2022},
  month = {Feb},
  volume = {602},
  pages = {245--250},
  doi = {10.1038/s41586-021-04327-z}
}

@article{nematicity.CsVTiSb.natphy,
  title = "{Unidirectional electron–phonon coupling in the nematic state of a kagome superconductor}",
  author = {Wu, Ping and Tu, Yubing and Wang, Zhuying and Yu, Shuikang and Li, Hongyu and Ma, Wanru and Liang, Zuowei  and Zhang, Yunmei and Zhang, Xuechen and Li, Zeyu and Yang, Ye and Qiao, Zhenhua and Ying, Jianjun and Wu, Tao and Shan, Lei and Xiang, Ziji and Wang, Zhenyu and Chen, Xianhui},
  journal = {Nature Physics},
  volume = {19},
  issue = {8},
  pages = {1143-1149},
  year = {2023},
  month = {Oct},
  doi = {10.1038/s41567-023-02031-5}
}

@article{lihong.natphy,
  title = "{Unidirectional coherent quasiparticles in the high-temperature rotational symmetry broken phase of AV$_3$Sb$_5$ kagome}",
  author = {Li, Hong and Zhao, He and Ortiz, Brenden R. and Oey, Yuzki and Wang, Ziqiang and Wilson, Stephen D. and Zeljkovic, Ilija},
  journal = {Nature Physics},
  volume = {19},
  issue = {5},
  pages = {637-643},
  year = {2023},
  month = {May},
  doi = {10.1038/s41567-022-01932-1}
}

@article{TiTeSe.nano.CDW,
  title = "{Controlling the Charge Density Wave Transition in Single-Layer TiTe_{2x}Se_{2(1–x)} Alloys by Band Gap Engineering}",
  author = {Antonelli, Tommaso and Rajan, Akhil and Watson, Matthew D. and Soltani, Shoresh and Houghton, Joe and Siemann, Gesa-Roxanne and Zivanovic, Andela and Bigi, Chiara and Edwards, Brendan and King, Phil D. C.},
  journal = {Nano Letters},
  volume = {24},
  issue = {1},
  pages = {215-221},
  year = {2024},
  month = {Jan},
  doi = {10.1021/acs.nanolett.3c03776}
}

@article{singleTiSe2.nc.CDW,
  title = "{Controlling the Charge Density Wave Transition in Single-Layer TiTe_{2x}Se_{2(1–x)} Alloys by Band Gap Engineering}",
  author = {Chen, P. and Chan, Y. -H. and Fang, X. -Y. and Zhang, Y. and Chou, M Y and Mo, S. -K. and Hussain, Z. and Fedorov, A. -V. and Chiang, T. -C.},
  journal = {Nature Communications},
  volume = {6},
  issue = {1},
  pages = {8943},
  year = {2015},
  month = {Nov},
  doi = {10.1038/ncomms9943}
}

@article{cxh2025.prx.CsV3Sb5-xSnx.stripe,
  title = "{Electron-Correlation-Assisted Charge Stripe Order in a Kagome Superconductor}",
  author = {Huai, Linwei and Wang, Zhuying and Rao, Huachen and Han, Yulei and Liu, Bo and Yu, Shuikang and Zhang, Yunmei and Zang, Ruiqing and Luan, Runqing and Peng, Shuting and Qiao, Zhenhua and Wang, Zhenyu and He, Junfeng and Wu, Tao and Chen, Xianhui},
  journal = {Phys. Rev. X},
  volume = {15},
  issue = {4},
  pages = {041039},
  numpages = {12},
  year = {2025},
  month = {Dec},
  publisher = {American Physical Society},
  doi = {10.1103/hfkr-k2pw},
}

@article{FeSeS.PNAS.2016.namtic,
author = {Suguru Hosoi  and Kohei Matsuura  and Kousuke Ishida  and Hao Wang  and Yuta Mizukami  and Tatsuya Watashige  and Shigeru Kasahara  and Yuji Matsuda  and Takasada Shibauchi },
title = "{Nematic quantum critical point without magnetism in FeSe$_{1-x}$S$_x$ superconductors}",
journal = {Proceedings of the National Academy of Sciences},
volume = {113},
number = {29},
pages = {8139-8143},
year = {2016},
doi = {10.1073/pnas.1605806113},
}

@article{E.Fradkin2015RMP,
  title = "{Colloquium: Theory of intertwined orders in high temperature superconductors}",
  author = {Fradkin, Eduardo and Kivelson, Steven A. and Tranquada, John M.},
  journal = {Rev. Mod. Phys.},
  volume = {87},
  issue = {2},
  pages = {457--482},
  numpages = {26},
  year = {2015},
  month = {May},
  publisher = {American Physical Society},
  doi = {10.1103/RevModPhys.87.457},
}

@article{ScV6Sn6.diffuse.scattering,
  title = "{Softening of a flat phonon mode in the kagome ScV$_6$Sn$_6$}",
  author={Korshunov, A. and Hu, H. and Subires, D. and Jiang, Y. and Călugăru, D. and Feng, X. and Rajapitamahuni, A. and Yi, C. and Roychowdhury, S. and Vergniory, M. G. and Strempfer, J. and Shekhar, C. and Vescovo, E. and Chernyshov, D. and Said, A. H. and Bosak, A. and Felser, C. and Bernevig, B. Andrei and Blanco-Canosa, S.},
  journal = {Nature Communications},
  volume = {14},
  issue = {1},
  pages = {6646},
  year = {2023},
  month = {Oct},
  doi = {10.1038/s41467-023-42186-6}
}

@article{sdH.prl.CVS,
  title = "{Quantum Transport Evidence of Topological Band Structures of Kagome Superconductor ${\mathrm{CsV}}_{3}{\mathrm{Sb}}_{5}$}",
  author = {Fu, Yang and Zhao, Ningning and Chen, Zheng and Yin, Qiangwei and Tu, Zhijun and Gong, Chunsheng and Xi, Chuanying and Zhu, Xiangde and Sun, Yuping and Liu, Kai and Lei, Hechang},
  journal = {Phys. Rev. Lett.},
  volume = {127},
  issue = {20},
  pages = {207002},
  numpages = {6},
  year = {2021},
  month = {Nov},
  publisher = {American Physical Society},
  doi = {10.1103/PhysRevLett.127.207002}
}

@article{HeZhao.nat.2021,
    title="{Cascade of correlated electron states in the kagome superconductor CsV$_3$Sb$_5$}",
    author={Zhao, He and Li, Hong and Ortiz, Brenden R. and Teicher, Samuel M. L. and Park, Takamori and Ye, Mengxing and Wang, Ziqiang and Balents, Leon and Wilson, Stephen D. and Zeljkovic, Ilija},
    volume={599},
    doi={10.1038/s41586-021-03946-w},
    number={7884},
    journal={Nature},
    publisher={Springer Science and Business Media LLC},
    year={2021},
    month={Sep},
    pages={216-221},
    language={en},
}

@article{mattieu.prl.pressure.cvs,
  title = "{Pressure-Dependent Electronic Superlattice in the Kagome Superconductor ${\mathrm{CsV}}_{3}{\mathrm{Sb}}_{5}$}",
  author = {Stier, F. and Haghighirad, A.-A. and Garbarino, G. and Mishra, S. and Stilkerich, N. and Chen, D. and Shekhar, C. and Lacmann, T. and Felser, C. and Ritschel, T. and Geck, J. and Le Tacon, M.},
  journal = {Phys. Rev. Lett.},
  volume = {133},
  issue = {23},
  pages = {236503},
  numpages = {7},
  year = {2024},
  month = {Dec},
  publisher = {American Physical Society},
  doi = {10.1103/PhysRevLett.133.236503},
}

@article{Nie.2022.nat,
title="{Charge-density-wave-driven electronic nematicity in a kagome superconductor}",
author={Nie, Linpeng and Sun, Kuanglv and Ma, Wanru and Song, Dianwu and Zheng, Lixuan and Liang, Zuowei and Wu, Ping and Yu, Fanghang and Li, Jian and Shan, Min and Zhao, Dan and Li, Shunjiao and Kang, Baolei and Wu, Zhimian and Zhou, Yanbing and Liu, Kai and Xiang, Ziji and Ying, Jianjun and Wang, Zhenyu and Wu, Tao and Chen, Xianhui},
journal={Nature}, 
volume={604},
year={2022}, 
month={Feb},
pages={59–64},
doi={10.1038/s41586-022-04493-8},
}

@article{F.H.Yu.pressure.nc,
  title = "{Unusual competition of superconductivity and charge-density-wave state in a compressed topological kagome metal}",
  author = {Yu, F. H. and Ma, D. H. and Zhuo, W. Z. and Liu, S. Q. and Wen, X. K. and Lei, B. and Ying, J. J. and Chen, X. H.},
  journal = {Nature Communications},
  volume = {12},
  issue = {1},
  pages = {3645},
  year = {2021},
  month = {Jun},
  doi = {10.1038/s41467-021-23928-w}
}

@article{XiangYing.nc.2fold.magnetoResi,
  title = "{Twofold symmetry of c-axis resistivity in topological kagome superconductor CsV$_3$Sb$_5$ with in-plane rotating magnetic field}",
  author={Xiang, Ying and Li, Qing and Li, Yongkai and Xie, Wei and Yang, Huan and Wang, Zhiwei and Yao, Yugui and Wen, Hai-Hu},
  journal = {Nature Communications},
  volume = {12},
  issue = {1},
  pages = {6727},
  year = {2021},
  month = {Nov},
  doi = {10.1038/s41467-021-27084-z}
}

@article{NLWang.TRS,
  title = "{Simultaneous formation of two-fold rotation symmetry with charge order in the kagome superconductor ${\mathrm{CsV}}_{3}{\mathrm{Sb}}_{5}$ by optical polarization rotation measurement}",
  author = {Wu, Qiong and Wang, Z. X. and Liu, Q. M. and Li, R. S. and Xu, S. X. and Yin, Q. W. and Gong, C. S. and Tu, Z. J. and Lei, H. C. and Dong, T. and others},
  journal = {Phys. Rev. B},
  volume = {106},
  issue = {20},
  pages = {205109},
  numpages = {7},
  year = {2022},
  month = {Nov},
  publisher = {American Physical Society},
  doi = {10.1103/PhysRevB.106.205109},
}

@article{LixuanZheng.nat.pressure.CVS,
  title = "{Emergent charge order in pressurized kagome superconductor CsV$_3$Sb$_5$}",
  author={Zheng, Lixuan and Wu, Zhimian and Yang, Ye and Nie, Linpeng and Shan, Min and Sun, Kuanglv and Song, Dianwu and Yu, Fanghang and Li, Jian and Zhao, Dan and Li, Shunjiao and Kang, Baolei and Zhou, Yanbing and Liu, Kai and Xiang, Ziji and Ying, Jianjun and Wang, Zhenyu and Wu, Tao and Chen, Xianhui},
  journal = {Nature},
  volume = {611},
  issue = {7927},
  pages = {682-687},
  year = {2022},
  month = {Dec},
  doi = {10.1038/s41586-022-05351-3}
}

@article{CVSSn.xrd.stripe,
  title = "{Incommensurate charge-stripe correlations in the kagome superconductor CsV$_3$Sb$_{5−x}$Sn$_x$}",
  author={Kautzsch, Linusand and Oey, Yuzki M. and Li, Hong and Ren, Zheng and Ortiz, Brenden R. and Pokharel, Ganesh and Seshadri, Ram and Ruff, Jacob and Kongruengkit, Terawit and Harter, John W. and Wang, Ziqiang and Zeljkovic, Ilija and Wilson, Stephen D.},
  journal = {npj Quantum Materials},
  volume = {8},
  issue = {1},
  pages = {37},
  year = {2023},
  month = {Jul},
  doi = {10.1038/s41535-023-00570-x}
}

@article{Lee.RMP.2006,
  title = "{Doping a Mott insulator: Physics of high-temperature superconductivity}",
  author = {Lee, Patrick A. and Nagaosa, Naoto and Wen, Xiao-Gang},
  journal = {Rev. Mod. Phys.},
  volume = {78},
  issue = {1},
  pages = {17--85},
  numpages = {0},
  year = {2006},
  month = {Jan},
  publisher = {American Physical Society},
  doi = {10.1103/RevModPhys.78.17},
}

@misc{2025.arxiv.persist.CDW,
      title="{Persistence of charge density wave fluctuations in the absence of long-range order in a hole-doped kagome metal}", 
      author={Terawit Kongruengkit and Andrea N. Capa Salinas and Ganesh Pokharel and Brenden R. Ortiz and Stephen D. Wilson and John W. Harter},
      eprint={2508.13290 (2025)},
      archivePrefix={arXiv},
}

@article{CVTiSb.PRL.2025.QPI,
  title = "{Revealing the Orbital Origins of Exotic Electronic States with Ti Substitution in Kagome Superconductor ${\mathrm{CsV}}_{3}{\mathrm{Sb}}_{5}$}",
  author = {Huang, Zihao and Chen, Hui and Tan, Hengxin and Han, Xianghe and Ye, Yuhan and Hu, Bin and Zhao, Zhen and Shen, Chengmin and Yang, Haitao and Yan, Binghai and Wang, Ziqiang and Liu, Feng and Gao, Hong-Jun},
  journal = {Phys. Rev. Lett.},
  volume = {134},
  issue = {5},
  pages = {056001},
  numpages = {7},
  year = {2025},
  month = {Feb},
  publisher = {American Physical Society},
  doi = {10.1103/PhysRevLett.134.056001},
}

@article{CVNS.ARPES.PRL,
  title = "{Fermiology and origin of ${T}_{c}$ enhancement in a Kagome superconductor $\mathrm{Cs}({\mathrm{V}}_{1\ensuremath{-}x}{\mathrm{Nb}}_{x}{)}_{3}{\mathrm{Sb}}_{5}$}",
  author = {Kato, Takemi and Li, Yongkai and Nakayama, Kosuke and Wang, Zhiwei and Souma, Seigo and Matsui, Fumihiko and Kitamura, Miho and Horiba, Koji and Kumigashira, Hiroshi and Takahashi, Takashi and others},
  journal = {Phys. Rev. Lett.},
  volume = {129},
  issue = {20},
  pages = {206402},
  numpages = {7},
  year = {2022},
  month = {Nov},
  publisher = {American Physical Society},
  doi = {10.1103/PhysRevLett.129.206402},
}

@article{Zhong.CVNS.CVTS.nat,
  title = "{Nodeless electron pairing in ${\mathrm{CsV}}_{3}{\mathrm{Sb}}_{5}$-derived kagome superconductors}",
  author = {Yigui Zhong and Jinjin Liu and Xianxin Wu and Zurab Guguchia and J.-X. Yin and Akifumi Mine and Yongkai Li and Sahand Najafzadeh and Debarchan Das and Charles Mielke III and others},
  journal = {Nature},
  volume = {617},
  pages = {488--492},
  year = {2023},
  month = {May},
  doi = {10.1038/s41586-023-05907-x},
}

@article{cpl2021RVS.sc,
author = {Qiangwei Yin and Zhijun Tu and Chunsheng Gong and Yang Fu and Shaohua Yan and Hechang Lei},
title = "{Superconductivity and Normal-State Properties of Kagome Metal ${\mathrm{RbV}}_{3}{\mathrm{Sb}}_{5}$ Single Crystals}",
publisher = {Chin. Phys. Lett.},
year = {2021},
journal = {Chinese Physics Letters},
volume = {38},
number = {3},
eid = {037403},
pages = {037403},
doi = {10.1088/0256-307X/38/3/037403}
}

@article{prm2019Brenden,
  title = "{New kagome prototype materials: discovery of ${\mathrm{KV}}_{3}{\mathrm{Sb}}_{5},{\mathrm{RbV}}_{3}{\mathrm{Sb}}_{5}$, and ${\mathrm{CsV}}_{3}{\mathrm{Sb}}_{5}$}",
  author = "{Ortiz, Brenden R. and Gomes, L\'{\i}dia C. and Morey, Jennifer R. and Winiarski, Michal and Bordelon, Mitchell and Mangum, John S. and Oswald, Iain W. H. and Rodriguez-Rivera, Jose A. and Neilson, James R. and Wilson, Stephen D. and Ertekin, Elif and McQueen, Tyrel M. and Toberer, Eric S.}",
  journal = {Phys. Rev. Mater.},
  volume = {3},
  issue = {9},
  pages = {094407},
  numpages = {9},
  year = {2019},
  month = {Sep},
  publisher = {American Physical Society},
  doi = {10.1103/PhysRevMaterials.3.094407},
  }

@article{BKeimer.review.nat.2015,
title = "{From quantum matter to high-temperature superconductivity in copper oxides}",
author = {Keimer, B. and Kivelson, S. A. and Norman, M. R. and Uchida, S. and Zaanen, J.},
journal = {Nature},
volume = {518},
issue = {7538},
pages = {179-186},
year = {2015},
month = {Feb},
doi = {10.1038/nature14165}
}

@article{Fernandes.nat,
title = "{What drives nematic order in iron-based superconductors?}",
author = {Fernandes, R. M. and Chubukov, A. V. and Schmalian, J.},
journal = {Nature Physics},
volume = {10},
issue = {2},
pages = {97-104},
year = {2014},
month = {Feb},
doi = {10.1038/nphys2877}
}

@article{MiaoQiSi.2016.nat.rev.mater,
title = "{High-temperature superconductivity in iron pnictides and chalcogenides}",
author = {Si, Qimiao and Yu, Rong and Abrahams, Elihu},
journal = {Nature Reviews Materials},
volume = {1},
issue = {4},
pages = {16017},
year = {2016},
month = {Mar},
doi = {10.1038/natrevmats.2016.17}
}

@article{Gegenwart.2008.natphy.heavyFermiMetal,
title = "{Quantum criticality in heavy-fermion metals}",
author = {Gegenwart, Philipp and Si, Qimiao and Steglich, Frank},
journal = {Nature Physics},
volume = {4},
issue = {3},
pages = {186--197},
year = {2008},
month = {Mar},
doi = {10.1038/nphys892}
}

@article{cuprate.RevModPhys.87.457,
  title = "{Colloquium: Theory of intertwined orders in high temperature superconductors}",
  author = {Fradkin, Eduardo and Kivelson, Steven A. and Tranquada, John M.},
  journal = {Rev. Mod. Phys.},
  volume = {87},
  issue = {2},
  pages = {457--482},
  numpages = {26},
  year = {2015},
  month = {May},
  publisher = {American Physical Society},
  doi = {10.1103/RevModPhys.87.457},
}

@article{Riccardo.2021.review.chargeOrder.cuprate,
author = {Arpaia ,Riccardo and Ghiringhelli ,Giacomo},
title = "{Charge Order at High Temperature in Cuprate Superconductors}",
journal = {Journal of the Physical Society of Japan},
volume = {90},
number = {11},
pages = {111005},
year = {2021},
doi = {10.7566/JPSJ.90.111005},
}

@article{PNAS.intertwined,
author = {J. C. Séamus Davis and Dung-Hai Lee},
title = "{Concepts relating magnetic interactions, intertwined electronic orders, and strongly correlated superconductivity}",
journal = {Proceedings of the National Academy of Sciences},
volume = {110},
number = {44},
pages = {17623-17630},
year = {2013},
doi = {10.1073/pnas.1316512110},
}

@article{AHE.CVS,
  title = {Concurrence of anomalous Hall effect and charge density wave in a superconducting topological kagome metal},
  author = {Yu, F. H. and Wu, T. and Wang, Z. Y. and Lei, B. and Zhuo, W. Z. and Ying, J. J. and Chen, X. H.},
  journal = {Phys. Rev. B},
  volume = {104},
  issue = {4},
  pages = {L041103},
  numpages = {7},
  year = {2021},
  month = {Jul},
  publisher = {American Physical Society},
  doi = {10.1103/PhysRevB.104.L041103}
}

@article{umR.kenny,
title = "{Absence of local moments in the kagome metal ${\mathrm{KV}}_{3}{\mathrm{Sb}}_{5}$ as determined by muon spin spectroscopy.}",
doi = {10.1088/1361-648X/abe8f9},
year = {2021},
month = {may},
publisher = {IOP Publishing},
volume = {33},
number = {23},
pages = {235801},
author = {Eric M Kenney and Brenden R Ortiz and Chennan Wang and Stephen D Wilson and Michael J Graf},
journal = {Journal of Physics: Condensed Matter},
}

\end{document}